\documentclass[10pt,conference]{IEEEtran}
\IEEEoverridecommandlockouts
\usepackage{cite}
\usepackage{amsmath,amssymb,amsfonts}
\usepackage{graphicx}
\usepackage{textcomp}
\usepackage{xcolor}
\usepackage{acronym}
\usepackage{multirow}
\usepackage{tabularx}
\usepackage{url}
\usepackage{algorithm}
\usepackage{algpseudocode}

\def\BibTeX{{\rm B\kern-.05em{\sc i\kern-.025em b}\kern-.08em
    T\kern-.1667em\lower.7ex\hbox{E}\kern-.125emX}}
\begin{document}

\title{Dithered Belief Propagation for Real-Time Quantum Memory Decoding}%

\author{
    \IEEEauthorblockN{Zhengyu Cai, Mohamad Ali Jarkas, Bokun Zhao, and Warren J. Gross}
    \IEEEauthorblockA{
        \textit{Department of Electrical \& Computer Engineering}\\
        \textit{McGill University}\\
        Montreal, QC, Canada\\
        \texttt{\{zhengyu.cai, mohamad.jarkas, bokun.zhao\}@mail.mcgill.ca, warren.gross@mcgill.ca}
    }
}

\maketitle

\begin{abstract}

We introduce dithered belief-propagation decoding for circuit-level quantum memory to mitigate the oscillations and trapping behavior that limit standard belief-propagation (BP) decoders for qLDPC codes. Our approach combines a quantum-adapted sign-flip mechanism with chained randomized prior perturbation. Evaluated under circuit-level noise on the distance-11 rotated surface code and the distance-12 bivariate-bicycle code, the proposed dithered decoder significantly outperforms BP and BP+OSD+CS-10. The proposed decoder achieves accuracy comparable to or better than state-of-the-art Relay-BP, especially at higher physical error rates, while trading a moderate increase in additions for about 98\% fewer multiplications on average and eliminating node-level RNGs. These results suggest that dithering is a promising approach toward real-time decoding of quantum memories.

\end{abstract}


\section{Introduction}

Fault-tolerant quantum computation relies on quantum error correction to protect logical information against the noise affecting physical qubits, gates, measurements, and idle operations\cite{steane1996error, hashim2025practical}. A fundamental requirement for fault tolerance is quantum memory: the continuous active preservation of an encoded logical state over multiple error-correction cycles\cite{yuan2021universal}. In practice, this protection requires a classical decoder that processes the syndrome data produced by repeated stabilizer measurements and outputs corrections quickly and accurately enough to keep pace with the quantum device\cite{battistel2023real}.

\begin{figure*}
    \centering
    \includegraphics[width = 0.8\textwidth]{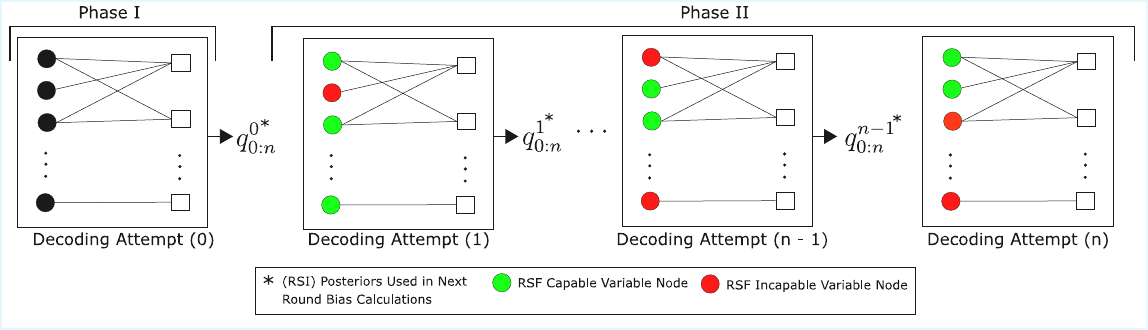}
    \caption{The proposed qDither Tanner graph architecture.}
    \label{fig:NetworkArchitecture}
\end{figure*}

Quantum LDPC codes have recently emerged as promising candidates for low-overhead fault tolerance\cite{bravyi2024high, kovalev2013quantum}, but their decoding remains difficult\cite{maan2026decoding}. Unlike simpler settings where some classical decoding ideas work well, qLDPC decoding graphs contain loops, symmetries, and degeneracies that cause standard belief propagation to oscillate, stall, and never converge, especially under circuit-level noise\cite{kuo2022exploiting, moradi2026learning}.

Minimum-weight perfect matching (MWPM) \cite{edmonds1965paths, higgott2022pymatching} is a natural reference point because it delivers high decoding accuracy for surface code\cite{bravyi1998quantum} architectures and has very fast CPU implementations. Its strength lies in exploiting the pairing structure of detection events, making it especially effective for topological codes\cite{breuckmann2016constructions, bonilla2021xzzx, li20192d} whose decoding graphs exhibit a well-aligned geometric structure. The limitation however is that this advantage does not transfer cleanly to general qLDPC circuit decoding due to the absence of suchstructure\cite{miao2025quaternary, higgott2022pymatching}.

For general qLDPC codes, a major breakthrough was the introduction of BP+OSD, where ordered statistics decoding post-processes the output of an non-convergent belief-propagation run using reliability information and Gaussian elimination \cite{panteleev2021degenerate, roffe2020decoding}. This approach was historically important because it enabled successful decoding in much of the qLDPC landscape and became a widely used gold-standard baseline. Its weakness is computational cost: OSD depends on expensive matrix operations, especially Gaussian elimination, which make it slow even on CPUs and particularly unattractive for low-latency real-time implementations \cite{valls2021syndrome, fan2025accelerating}. Therefore, BP+OSD should be framed as a strong accuracy-oriented baseline, but one whose heavy post-processing motivates the search for decoders without needing Gaussian elimination while retaining comparable performance.

There had already been substantial effort to suppress trapping-set and oscillation pathologies while staying within a message-passing framework. Classical dithered BP\cite{leduc2012dithered} is an early example: it introduces randomized perturbations such as random sign flips and randomized initial states to escape BP failure modes, and was explicitly designed to reduce error floors with minimal hardware overhead rather than through OSD-like post-processing. In the qLDPC setting, work on enhanced Min-Sum~\cite{chytas2025enhanced} considered the oscillatory behavior of Min-Sum on symmetric stabilizers of two-block CSS codes and proposed an iteration-dynamics-memory variant in which past messages are reused and, when consecutive variable-to-check messages disagree in sign, the current and previous messages are combined to dampen the oscillation. They report large gains over normalized Min-Sum and even slight improvements over BP+OSD-0\cite{roffe2020decoding}, i.e., OSD with no combination sweep, on binary-symmetric-channel bivariate-bicycle benchmarks while retaining linear complexity. However, these efforts still did not establish a realistic circuit-level alternative to BP+OSD for general qLDPC decoding.

Recent work\cite{muller2025improved, maurer2025real} by IBM has shown that one does not need to rely on OSD-style post-processing to exceed BP+OSD-level decoding performance. Relay-BP demonstrates this by augmenting belief propagation with disordered memory strengths and ensembling, achieving substantially better performance than BP+OSD+CS-10 on bivariate-bicycle benchmarks\cite{kovalev2013quantum}. As a result, Relay-BP provides a natural modern baseline for message-passing-based qLDPC quantum-memory decoding \cite{maurya2025fpga}. Instead of using a fixed prior term throughout the iterations, Relay-BP recomputes a memory-weighted bias at every timestep. This extra repeated per-iteration update makes Relay-BP structurally more involved and diverges from simple, hardware-friendly primitives that Min-Sum provides.

In this work, we introduce an enhanced quantum variant of the dithered Min-Sum decoder that conquers non-converging behavior of belief propagation while retaining the simplest message-passing primitives. 

Our contributions are listed as follows:

\begin{itemize}
    \item We propose qDither, a quantum dithered min-sum decoding algorithm that outperforms Relay-BP on surface code and bivariate-bicycle code (Gross code) at higher noise levels.
    \item We show that qDither eliminates 98\% of the multiplications in Relay-BP at the cost of a moderate increase in additions, and requires minimal augmentation to the standard syndrome-based Min-Sum message-passing structure making it more suitable for real-time decoding implementation.
\end{itemize}

\section{Background}
Quantum error correction protects logical information by repeatedly measuring check operators and using the resulting classical syndrome data to infer a correction. In the idealized picture, an unknown physical error acts on the encoded state, the measured syndrome identifies the violated checks, and a decoder returns a correction that aims to restore the state to the codespace\cite{KissingerWetering2024Book, kuo2022exploiting, devitt2013quantum}.

A decoding instance is specified by a parity-check matrix and a logical observable matrix
\[
H \in \mathbb{F}_2^{M \times N},\hspace{0.5cm}
A \in \mathbb{F}_2^{K \times N},
\]
and a fault-probability vector
\[
p = (p_1,\dots,p_N) \in [0,1]^N.
\]
Here, the \(N\) columns correspond to elementary error events, or fault locations, and the \(M\) rows correspond to detector outcomes.
Let
\[
E \in \mathbb{F}_2^N
\]
denote the unknown error pattern, where \(E_j = 1\) indicates that error event \(j\) occurred. Assuming independent error events, the probability of \(E\) is
\begin{equation}
\Pr(E)
=
\prod_{j=1}^{N}
(1-p_j)
\left(\frac{p_j}{1-p_j}\right)^{E_j}.
\label{eq:error-probability}
\end{equation}
The observed syndrome is the binary vector
\begin{equation}
\sigma = HE \in \mathbb{F}_2^M,
\label{eq:syndrome-definition}
\end{equation}
where all arithmetic is over \(\mathbb{F}_2\). The logical action induced by the error pattern is similarly represented by
\begin{equation}
a(E) = AE \in \mathbb{F}_2^K.
\label{eq:logical-action}
\end{equation}
A decoder is a classical map
\[
D : \mathbb{F}_2^M \to \mathbb{F}_2^N
\]
that takes the syndrome \(\sigma\) as input and outputs an estimated correction
\[
\hat E = D(\sigma).
\]
The decoder should return a syndrome-consistent estimate, i.e.
\begin{equation}
H\hat E = \sigma.
\label{eq:syndrome-consistency}
\end{equation}
However, the true objective is not exact recovery of \(E\), since many distinct error patterns can produce the same syndrome. Instead, decoding succeeds when the residual error \(E+\hat E\) has trivial logical action:
\begin{equation}
A(E+\hat E)=0.
\label{eq:decoding-success}
\end{equation}
Equivalently, \(\hat E\) and \(E\) must belong to the same logical coset.

This binary description induces a Tanner graph \(G(H)\). The graph is bipartite: each column of \(H\) defines an variable node, each row defines a check node, and an edge connects variable node \(j\) to check node \(i\) whenever \(H_{ij}=1\). We write \(N(i)\) for the set of variable nodes adjacent to check node \(i\), and \(N(j)\) for the set of check nodes adjacent to variable node \(j\). In this graph-theoretic picture, the syndrome is the set of check nodes incident to an odd number of active variable nodes, and decoding amounts to finding a set of variable nodes whose boundary reproduces the observed syndrome.

\subsection{Syndrome based Min-Sum Decoding}
\label{subsec:minsum}

We next recall the syndrome-based Min-Sum decoder\cite{fossorier1999reduced} for the binary quantum decoding problem. The decoder operates on the Tanner graph associated with the binary check matrix \(H\), where each column of \(H\) corresponds to an variable node and each row corresponds to a check node. Let \(N(i)\) denote the set of variable nodes adjacent to check node \(i\), and let \(N(j)\) denote the set of check nodes adjacent to variable node \(j\).

For each error location \(j\), we define the prior log-likelihood ratio
\begin{equation}
\lambda_j = \ln\!\left(\frac{1-p_j}{p_j}\right).
\label{eq:prior-llr}
\end{equation}
In standard quantum Min-Sum decoding, these prior terms remain fixed throughout the iterations.

The decoder iteratively exchanges real-valued messages along edges of the Tanner graph. We denote by \(m_{j \to i}^{(\ell)}\) the message sent from variable node \(j\) to check node \(i\) at iteration \(\ell\), and by \(m_{i \to j}^{(\ell)}\) the message sent in the reverse direction.

\paragraph{Initialization.}
At the start of decoding, each variable-to-check message is initialized by the corresponding prior:
\begin{equation}
m_{j \to i}^{(0)} = \lambda_j .
\label{eq:minsum-init}
\end{equation}

\paragraph{Check-node update.}
Given the incoming variable-to-check messages from the previous iteration, the check-to-variable update is

\begin{equation}
s_{i \to j}^{(\ell)}
=
(-1)^{\sigma_i}
\prod_{j' \in N(i)\setminus\{j\}}
\operatorname{sgn}\!\bigl(m_{j' \to i}^{(\ell-1)}\bigr)
\label{eq:temp-sign}
\end{equation}

\begin{equation}
m_{i \to j}^{(\ell)}
=
s_{i \to j}^{(\ell)}
\min_{j' \in N(i)\setminus\{j\}}
\left| m_{j' \to i}^{(\ell-1)} \right|
\label{eq:minsum-check}
\end{equation}

\paragraph{Variable-node update.}
Each variable node then combines its prior with the incoming check messages from all neighboring checks except the destination check:
\begin{equation}
m_{j \to i}^{(\ell)}
=
\lambda_j
+
\sum_{i' \in N(j)\setminus\{i\}}
m_{i' \to j}^{(\ell)} .
\label{eq:minsum-variable}
\end{equation}

\paragraph{Marginals and hard decisions.}
After the edge updates at iteration \(\ell\), the marginal log-likelihood ratio at variable node \(j\) is computed as
\begin{equation}
L_j^{(\ell)}
=
\lambda_j
+
\sum_{i' \in N(j)}
m_{i' \to j}^{(\ell)} .
\label{eq:minsum-marginal}
\end{equation}
Negative values of \(L_j^{(\ell)}\) favor the presence of an error. Accordingly, the hard decision is
\begin{equation}
\hat E_j^{(\ell)}
=
\begin{cases}
1, & L_j^{(\ell)} < 0,\\
0, & L_j^{(\ell)} \ge 0.
\end{cases}
\label{eq:minsum-hard}
\end{equation}

\paragraph{Stopping rule.}
Let
\[
\hat E^{(\ell)} =
\bigl(\hat E_1^{(\ell)},\hat E_2^{(\ell)},\dots,\hat E_N^{(\ell)}\bigr)
\in \mathbb{F}_2^N .
\]
If the estimated error pattern satisfies
\begin{equation}
H \hat E^{(\ell)} = \sigma ,
\label{eq:minsum-valid}
\end{equation}
Then the decoder halts and outputs \(\hat E^{(\ell)}\); otherwise, the iterations continue until a pre-defined maximum iteration budget is reached.

\subsection{Classical Dithered Min-Sum Decoding}
In a dithered decoder, in the event that the decoder fails to find a solution, a randomized second decoding phase (Phase-II) is invoked\cite{leduc2012dithered}. The two dithering mechanisms are:

\medskip
\noindent
\paragraph{Random sign flip (RSF)}
In this variant, the Min-Sum check-node update is modified by a random sign variable. For each check node \(i\) and iteration \(\ell\), let
\[
a_i^{(\ell)} \in \{-1,+1\},
\qquad
\Pr\!\bigl(a_i^{(\ell)}=-1\bigr)=\rho,
\]
with all draws independent. The check-to-variable message becomes
\begin{equation}
m{'}_{i\to j}^{(\ell)}
=
a_i^{(\ell)}\,m_{i\to j}^{(\ell)}
\end{equation}

\medskip
\noindent

\paragraph{Random initial state (RIS)}
In this variant, one performs ensembling multiple decoding trials with randomly perturbed inputs. For trial \(r\), define a dither vector
\[
d^{(r)}=(d_1^{(r)},\dots,d_N^{(r)}),
\]
whose components are i.i.d. with zero mean. The \(r\)-th trial then runs the ordinary Min-Sum recursion with initialization

\begin{equation}
\lambda_j^{(r)}=\lambda_j+d_j^{(r)}=m_{j\to i}^{(r,0)}.
\label{eq:ris-init}
\end{equation}

\section{Quantum Dithered Min-Sum Decoding (\textnormal{q}Dither)}

In this quantum variant of dithered Min-Sum (qDither), as shown in Figure~\ref{fig:NetworkArchitecture}, we modify both RSF and RIS to better address degeneracy-induced oscillations.

\paragraph{Enhanced RSF}
We build upon the soft sign-flip rule proposed by Chytas \emph{et al.}\cite{chytas2025enhanced}. Utilizing the symmetric structure of CSS stabilizers, their method relies on separating the XZ decoding problem into two identical sub-decoding problems at code-level. We extend this approach into a stochastic variant to serve as a true dithering mechanism for general qLDPC codes at circuit-level. Instead of randomly flipping check-node signs like classical RSF, we first randomly partition all variable nodes into two groups. The first group performs the standard variable node update, while the second group detect sign flips between the current and previous message and merges them.

\begin{equation}
\delta_{j \to i}^{(\ell)} =
\frac{1 - \operatorname{sgn}\!\left(m_{j \to i}^{(\ell)}\right)\operatorname{sgn}\!\left(m_{j \to i}^{(\ell-1)}\right)}{2}
\label{eq:eRsf}
\end{equation}
\noindent
\begin{equation}
m_{j\to i}^{\prime\prime(\ell)}
=
\delta_{j\to i}^{(\ell)}\,m_{j\to i}^{(\ell-1)}
+
m_{j\to i}^{(\ell)}
\label{eq:eRis}
\end{equation}

The probability \(\rho\) that a variable node receives this mechanism is a hyperparameter. Larger \(\rho\) increases the influence of the soft sign-flip correction rule, which may stabilize oscillatory message dynamics and lead to faster convergence, but can also reduce exploratory diversity. This mechanism may also be enabled in Phase I.

\medskip
\noindent

\paragraph{Enhanced RIS}
Instead of ensembling independent trials, Phase-II now employs a chained design in which the decoder inherits the posterior produced by the preceding decoder, to form a new input bias by blending it with the fixed base bias.

For each variable node \(j\), the Phase-II bias for the \(r\)th decoder is defined as
\begin{equation}
\lambda_j^{(r)}
=
\bigl(1-\chi_j^{(r)}\bigr)\lambda_j^{\mathrm{base}}
+
\chi_j^{(r)} q_j^{(r-1)},
\qquad
\chi_j^{(r)} \sim \mathrm{Unif}[\alpha,\beta]
\label{eq:chained_ris}
\end{equation}
where $q^{(r-1)}$ is the posterior of $(r-1)$th decoder.

The parameters \(\alpha\) and \(\beta\) bound the random mixing strength \(\chi\). In our experiments, we use \(\beta \leq 1.0\) and \(\alpha \ge -0.15\) as conservative settings, so that inherited posterior information influences the base prior, while still allowing occasional negative correction to escape an unfavorable inherited direction.

The pseudocode of the proposed qDither procedure is summarized in Algorithm 1.

\renewcommand{\algorithmicrequire}{\textbf{Input:}}
\renewcommand{\algorithmicensure}{\textbf{Output:}}

\begin{algorithm}
\caption{Quantum Dithered Min-Sum}
\begin{algorithmic}[1]
\Require $H, \sigma, p, n, b, \alpha, \beta, \rho$
\Ensure $\hat{E}$
\Statex\textbf{Phase I:}
\State Compute $\lambda_j^{base}$ from $p$ \hfill // Eq.~(6)
\State $m_{j\rightarrow i}^{(0)} \leftarrow \lambda_j^{base}$ \hfill // Eq.~(7)
\State Compute $\hat{E}$ via standard Min-Sum \hfill // Eqs.~(8)--(11)
\If{$H\hat{E} = \sigma$} 
    \State \Return $\hat{E}$
\EndIf

\Statex\textbf{Phase II:}
\State $q_j^{(0)} \leftarrow \lambda_j^{base}$

\For{$r = 1 \dots n$}
    \State Draw $\chi_j^{(r)} \sim \text{Unif}[\alpha, \beta]$
    \State Compute $\lambda_j^{(r)}$ from $\lambda_j^{base}, \chi_j^{(r)}, q_j^{(r-1)}$ \hfill // Eq.~(18)
    \State $m_{j\rightarrow i}^{(0)} \leftarrow \lambda_j^{(r)}$
    
    \State Draw $I_j^{(r)} \sim \text{Bernoulli}(\rho)$ 
    
    \For{$l = 1 \dots b$}
        \State Compute $m_{i\rightarrow j}^{(l)}$ \hfill // Eq.~(9)

        \State Compute $m_{j\rightarrow i}^{(l)}$ \hfill // Eq.~(10)
        
        \If{$I_j^{(r)} == 1$}
        \If{sgn$(m_{j\rightarrow i}^{(l)})\neq$ sgn$(m_{j\rightarrow i}^{(l-1)})$}
             \State $m_{j\rightarrow i}^{(l)}$ += $m_{j\rightarrow i}^{(l-1)}$
        \EndIf\EndIf

        \State Compute $L_j^{(l)}$ from $\lambda_j^{(r)}, m_{i\rightarrow j}^{(l)}$ \hfill // Eq.~(11)
        \State Compute $\hat{E}^{(l)}$ from $L_j^{(l)}$ \hfill // Eq.~(12)
        
        \If{$H\hat{E}^{(l)} = \sigma$}
            \State \Return $\hat{E}^{(l)}$
        \EndIf
    \EndFor
    
    \State $q_j^{(r)} \leftarrow L_j^{(l)}$
\EndFor

\State \Return $\hat{E}^{(l)}$
\end{algorithmic}
\end{algorithm}

\section{Results} \label{sec:Results}

We benchmark the proposed decoder under a circuit-level noise model\cite{bravyi2024high, bravyi1998quantum} using the same evaluation protocol adopted in prior circuit-based BP studies\cite{muller2025improved, maurer2025real, maan2026decoding, maurya2025fpga}. Each benchmark instance is generated by repeating noisy measurements and appending a final noiseless round. Under the circuit-level noise model, each elementary circuit location fails independently according to its corresponding fault channel, and each individual fault mechanism is treated as a candidate error event in the decoding problem\cite{perlin2023qldpc, gidney2021stim}.

These fault events induce a decoding object specified by a circuit-level parity-check matrix \(H\), a logical observable matrix \(A\), and a prior probability vector \(p\). The matrix \(H\) records which detector outcomes are flipped by each fault event, while \(A\) records whether those faults flip the logical observables of the code. The decoder is then supplied only with the observed detector vector and must output a candidate correction consistent with this induced decoding instance.

To reduce redundancy, fault events with identical detector signatures may be compressed into a single effective variable with a new prior. Decoder performance is evaluated by Monte Carlo simulation over a range of physical noise rates. Both of our decoding examples used are CSS-type\cite{calderbank1996good}. We will evaluate primarily the logical error rate (LER) and secondly, the iteration count. The hyperparameters are chosen and optimized through sweeping over 1000 samples at noise level $p=0.005$.

\subsection{Surface Code}
We consider the circuit-level decoding problem generated by the distance-11 rotated surface code\cite{horsman2012surface}. We derive separate \(H_X\) and \(H_Z\) decoding instances through basis filtering and perform XZ decoding. We use \(3000\) decoder chains, draw the dither as \(\chi \sim \operatorname{Unif}[-0.14,\,0.85]\), and limit each decoder call to at most \(b=15\) iterations. In addition, we allow the decoder to collect up to five distinct solutions and return the one with the lowest weighted LLR.

\begin{figure}
    \centering
    \includegraphics[width = 0.4\textwidth]{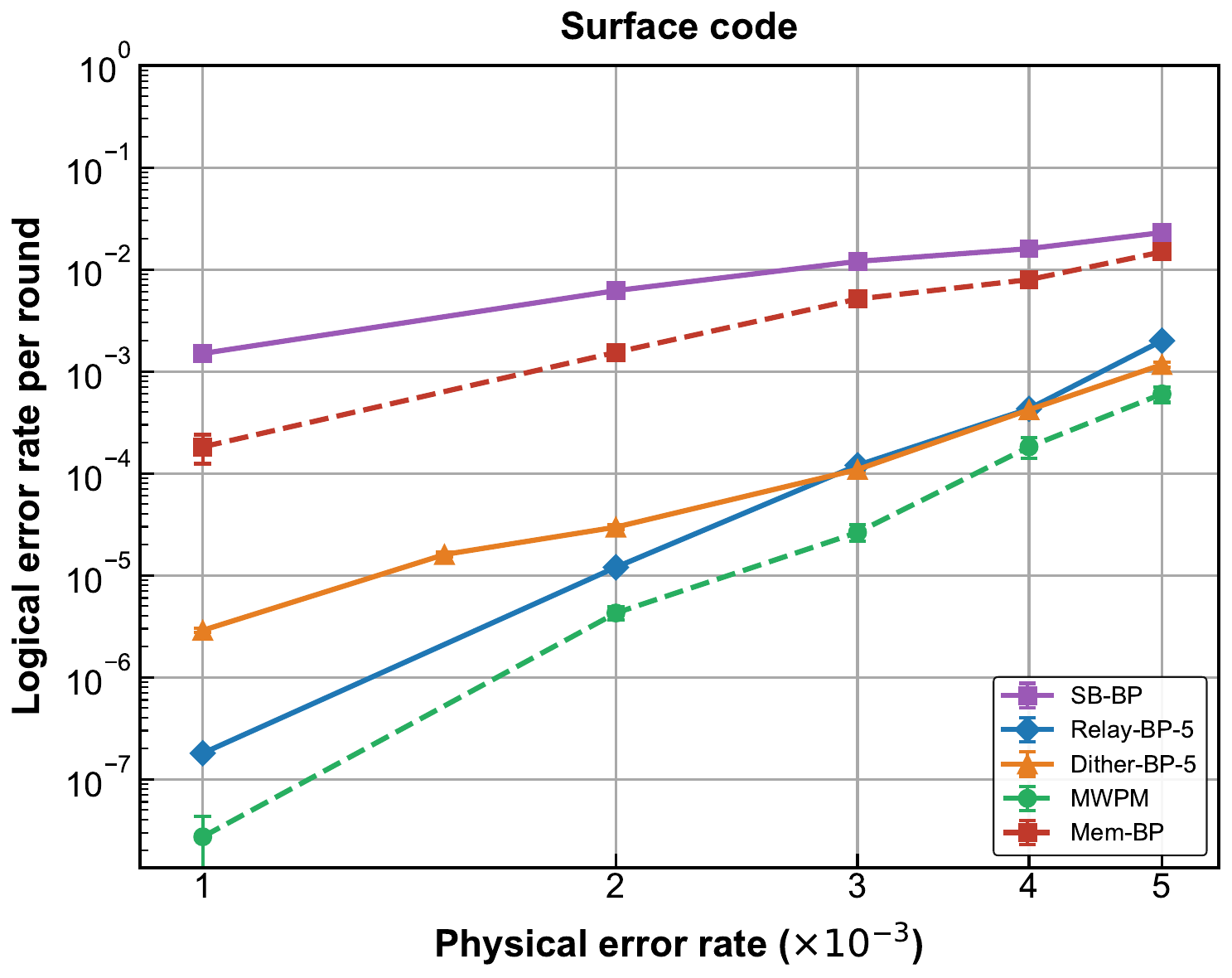}
    \caption{LER of decoding distance-11 rotated surface code, compared with syndrome-based Min-Sum (SB-BP), Relay-BP, Mem-BP and MWPM.}
    \label{fig:Surface}
\end{figure}

Figure~\ref{fig:Surface} shows the logical error rate per cycle---equivalently, per round in this setting---for varying noise levels and different decoders. For \(p \ge 0.003\), qDither-BP-5 outperforms Relay-BP-5\cite{muller2025improved}. Although MWPM\cite{higgott2022pymatching} still achieves the lowest LER, qDither-BP surpasses the remaining message-passing decoders by orders of magnitude. Mem-BP\cite{chen2025improved, nachmani2018deep, muller2025improved} introduces a memory term that updates a local bias every iteration. Relay-BP builds upon Mem-BP by randomizing the memory strength across relay legs and ensembling decoders by passing forward the final marginals. Structurally, qDither-BP is a low-overhead perturbative extension of standard Min-Sum-BP, whereas Relay-BP is a stateful multi-leg decoder that augments BP with dynamic memory updates.

\subsection{Gross code}
Next, we test our decoder on the \([[144,12,12]]\) bivariate-bicycle code, also known as the Gross code\cite{bravyi2024high}. The full parity-check matrix has shape \(1872 \times 71280\). In this experiment, we examine both XZ and XYZ decoding, with RSF enabled for both phases. In XYZ decoding, we pass the full parity-check matrix directly into the decoder, thereby preserving correlations between errors. In the memory benchmark, 12 Bell pairs are prepared, with one half of each pair fed into the Gross-code encoder while the other halves remain as noiseless references. Here we use \(1500\) decoder chains, a decoding budget of \(b=100\), a dither range \(\chi \sim \operatorname{Unif}[-0.07,\,0.5]\), and \(\rho=0.65\).

\begin{figure}
    \centering
    \includegraphics[width = 0.4\textwidth]{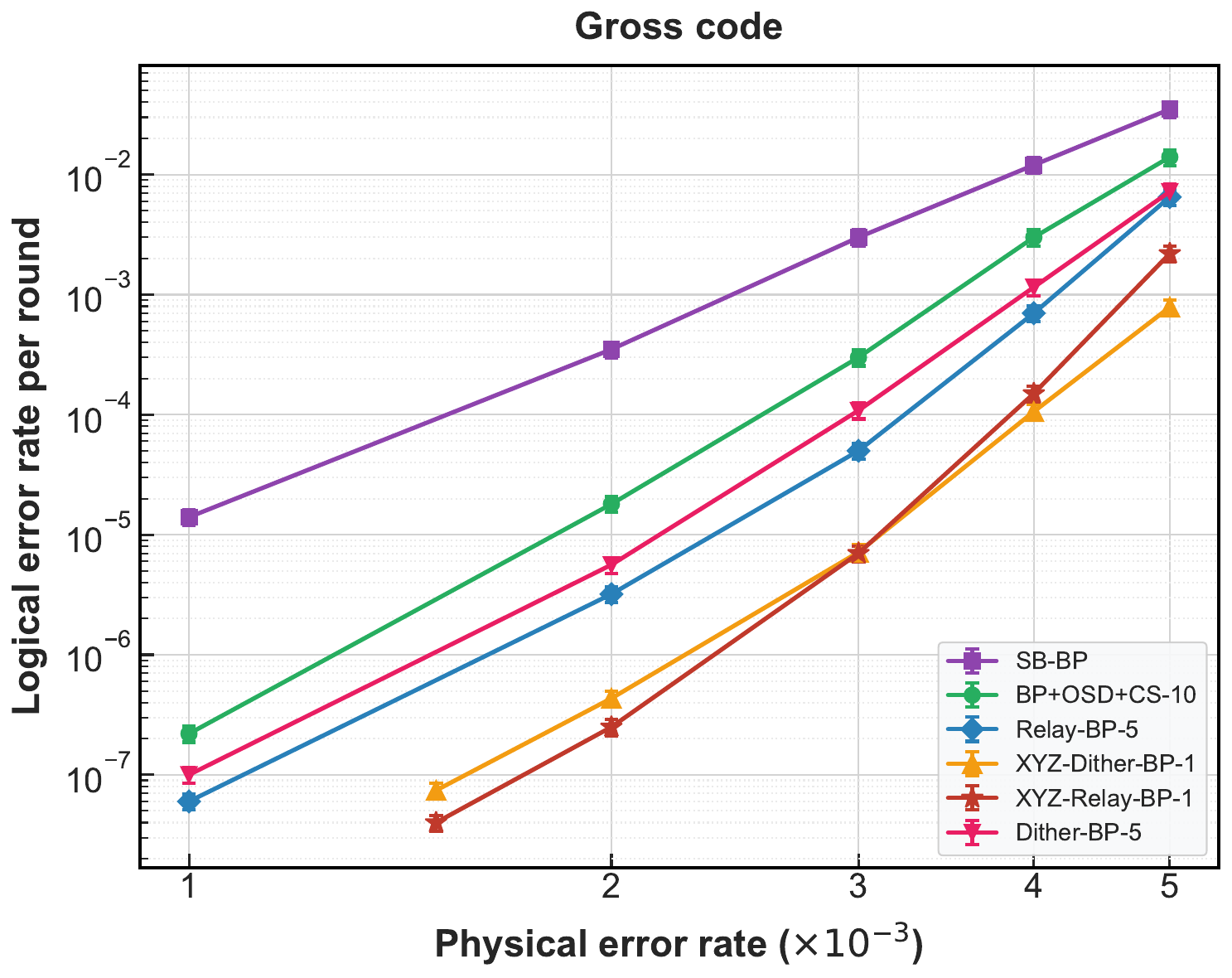}
    \caption{LER of decoding Gross code, compared with syndrome-based Min-Sum (SB-BP), BP+OSD-CS-10, Relay-BP-5, qDither-BP-5 with XZ decoding, as well as XYZ-Relay-BP-1 and XYZ-qDither-BP-1.}
    \label{fig:gross}
\end{figure}

Figure~\ref{fig:gross} evaluates our decoders on the Gross code. For independent XZ decoding, we benchmark against syndrome-based Min-Sum, Relay-BP-5 and BP+OSD+CS-10\cite{roffe2020decoding} that allows OSD to use up to 10 combination sweeps for syndrome matching. The overall trend is that plain syndrome-based Min-Sum remains the weakest method, surpassed by other models by a huge margin. Notably, qDither-BP-5 demonstrates accuracy nearly on par with Relay-BP-5, and outperforms BP+OSD+CS-10 at all noise levels.

For XYZ decoding, we compare XYZ-qDither-1 with XYZ-Relay-1. At \(p=3\times 10^{-3}\), the proposed XYZ-qDither-BP-1 achieves a logical error rate of approximately \((7.15\pm0.5)\times 10^{-6}\), which is very competitive with the \((7\pm1)\times10^{-6}\) reported for XYZ-Relay-1. At \(p=5\times 10^{-3}\), notably, qDither-BP outperforms Relay-BP, suggesting that dithering provides a stronger advantage in the higher-noise regime. This noise range reflects the noise levels of near-term superconducting quantum architectures\cite{maldonado2022error, georgopoulos2021modeling, google2023suppressing}, and it aligns exactly with the benchmarking regime established in prior works\cite{chen2025improved, maurer2025real, muller2025improved}. Taken together, these results indicate that qDither-BP attains Relay-level performance at low physical error rates while offering improved robustness at larger \(p\).

\section{Real-time decoding}
\subsection{Algorithmic Complexity}
To achieve real-time decoding, the decoder must produce results faster than the generation of a QEC round to avoid the buffer backlog issue\cite{Terhal_2015}. For the distance-12 bivariate-bicycle code, using a \(1\,\mu\mathrm{s}\) QEC cycle time\cite{battistel2023real} would allow a temporal budget of \(12\,\mu\mathrm{s}\) per window. In order to meet this timing requirement, Müller \emph{et al.}\cite{muller2025improved} suggest that an unrolled design with a critical path that allows for a clock period of 20 ns (per iteration) would translate to a $\sim600$ BP iteration budget. Consequently, a decoder is only practically viable in this regime if it achieves low logical error rates while keeping its average iteration count within this effective real-time budget, assuming a comparable per-iteration workload.

In this work, we consider algorithmic complexity from two complementary perspectives, namely iteration and operation complexity. The former measures the average number of BP iterations needed to perform one decoding task; the latter accounts for the number of operations within each BP iteration.

\subsubsection{Iteration Complexity}

\begin{figure}[!t]
    \centering
    \includegraphics[width = 0.5\textwidth]{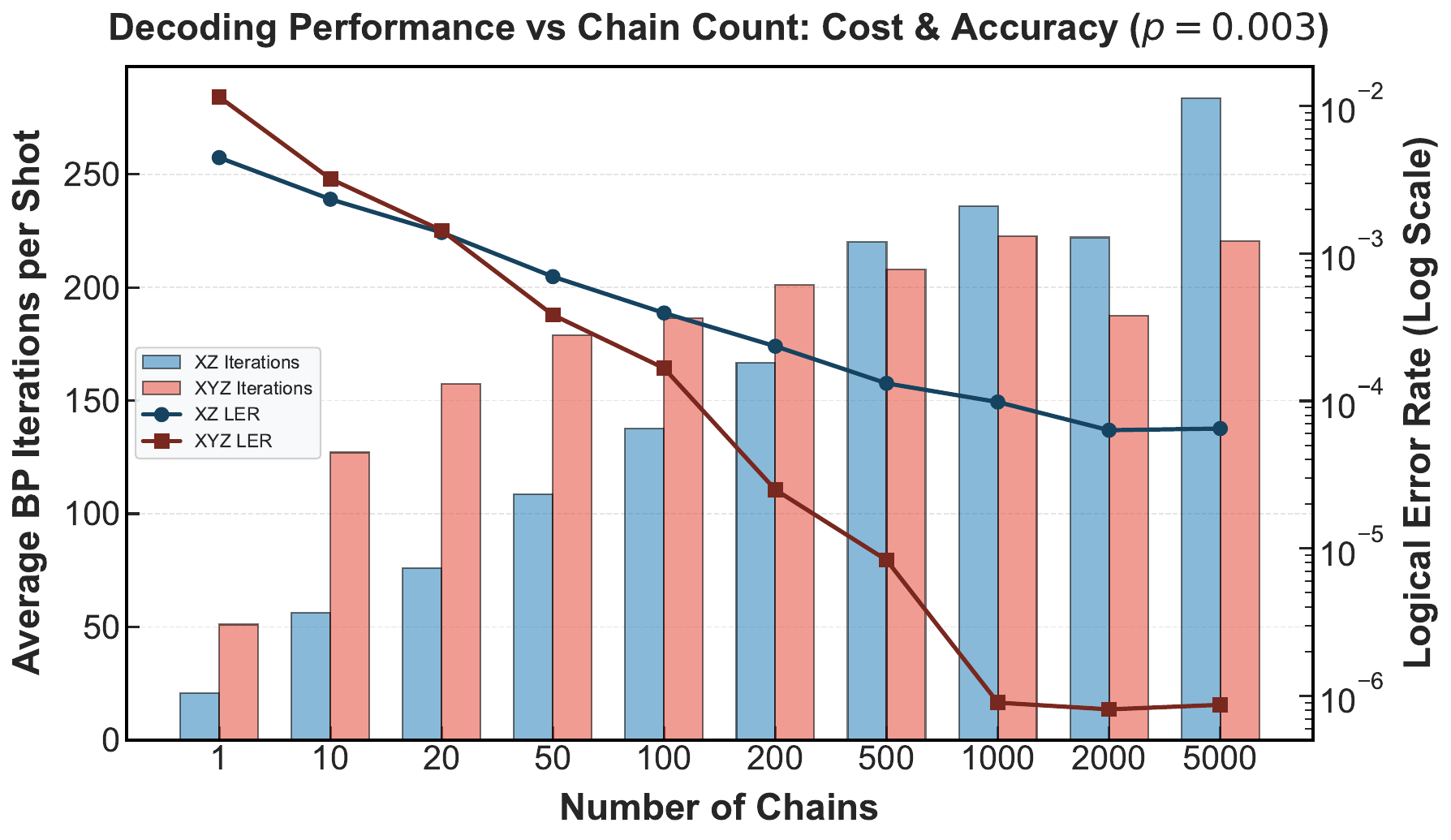}
    \caption{Average BP iteration per shot for decoding Gross code with varying number of chains and respective LER at noise level $p=0.003$.}
    \label{fig:chains}
\end{figure}

Figure~\ref{fig:chains} shows a clear tradeoff between decoder cost and accuracy as the number of chains increases at \(p=0.003\). For both XZ and XYZ decoding with 1 solution sought, increasing the chain count steadily lowers the logical error rate, with the largest gains occurring at small to intermediate chain counts. With more chains, however, the improvement in logical error rate becomes more gradual and the average iteration count appears to cap under 250 iterations per shot despite having more available budget. This saturation is particularly noticeable for XYZ-qDither. This shows that even though we use a high number of chained decoders, qDither manages to decode well under the required iteration budget.

\begin{figure}[!t]
    \centering
    \includegraphics[width = 0.4\textwidth]{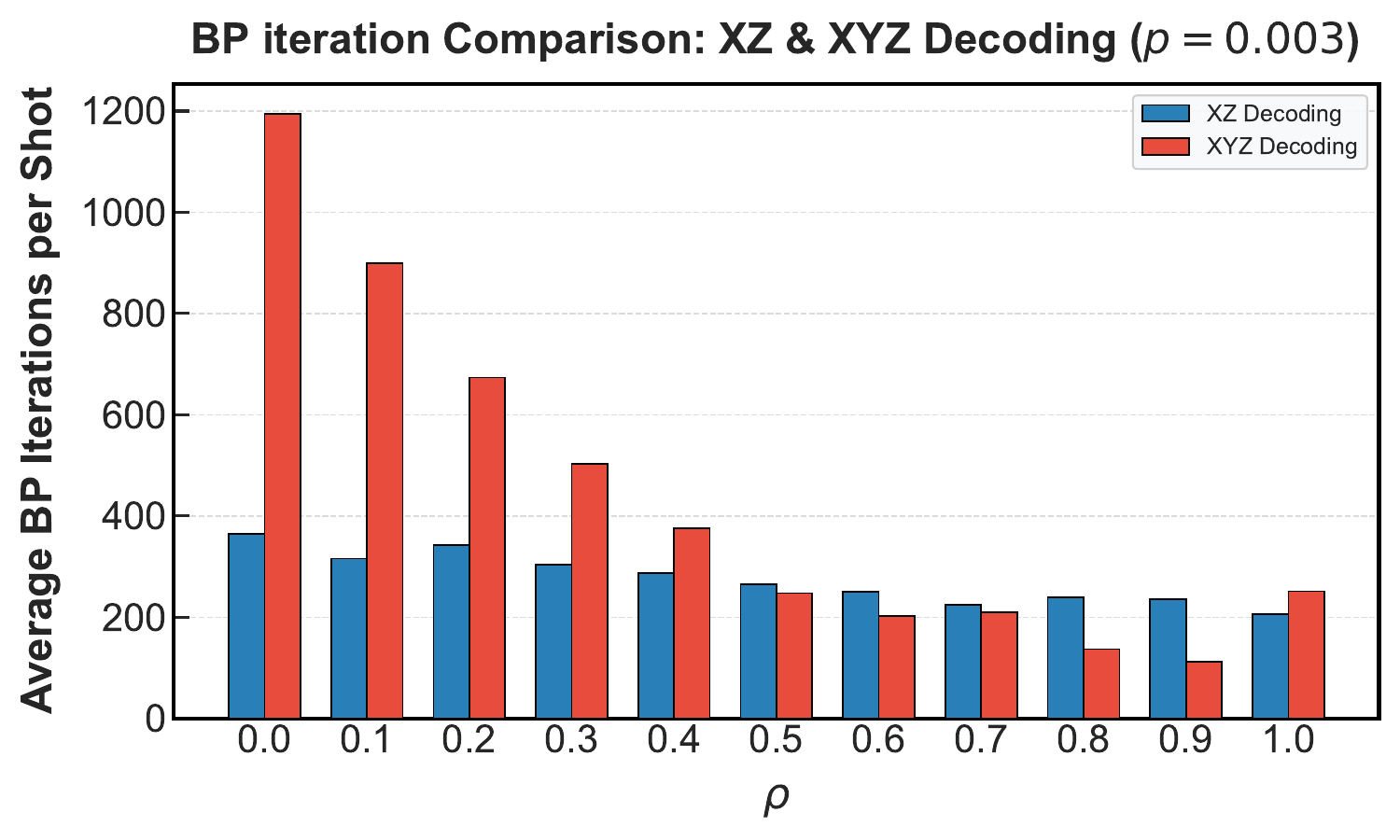}
    \caption{Average BP iteration per shot for decoding Gross code under varying RSF $\rho$ at noise level $p=0.003$.}
    \label{fig:iterations_comparison}
\end{figure}

Figure~\ref{fig:iterations_comparison} shows how the average decoding cost changes as the RSF parameter \(\rho\) is varied at \(p=0.003\). When RSF is disabled, i.e., standard Syndrome-based Min-Sum, the decoder requires a large number of BP iterations per shot, making it hard to satisfy a realistic real-time decoding budget; this effect is especially pronounced for XYZ decoding. As \(\rho\) is increased from zero, the average iteration count drops sharply. Interestingly, the cost rises again near \(\rho=1\), suggesting that this extra degree of perturbation helps dampen oscillation and our proposed RSF mechanism provides substantial speedup in iterations needed to converge.

\subsubsection{Operation Complexity}
Let $\xi, V, N, t$ be the number of edges, variable nodes, average chains and iterations respectively. A variable node gathers all incoming messages and then subtracts exactly one of them for each outgoing message; this corresponds to a total of $2\xi$ number of additions. Internal RSF yields approximately $\rho \xi$ extra additions. Relay-BP on the other hand, requires $2\xi + V$ additions and $2V$ multiplications for node-wise bias update.

To set up each Phase-II RIS, one computes new priors which demands a total of $2(N-1)V$ multiplications and $(N-1)V$ additions. The XYZ parity-check matrix for the Gross code has shape \(1872 \times 71{,}280\), with \(\xi = 419{,}904\). At noise level \(p = 0.003\), XYZ-qDither-1 has \(V = 71{,}280\), \(N = 2.6\), and \(t = 200\). This yields a setup cost of \(228{,}096\) multiplications and \(114{,}048\) additions, as well as a computation cost of \(530\xi\), for a total of \(222{,}663{,}168\) additions. We implemented XYZ-Relay-1 using the best parameters reported. This model reports \(t = 85\), yielding a total cost of \(77{,}442{,}480\) additions and \(12{,}117{,}600\) multiplications.

These figures indicate that an unrolled hardware implementations of Relay-BP would be substantially larger. Although qDither requires approximately \(2.88\times\) more additions, it uses only \(2\%\) of the multiplications required by Relay-BP. This gives qDither a clear advantage due to the complete absence of multiplication during effective decoding stage.

\subsection{Hardware Mapping}
Valls \emph{et al.}\cite{valls2021syndrome} suggest an unrolled architecture for VNUs and CNUs to achieve the latency required to avoid the backlog issue. Recently, Maurer \emph{et al.}\cite{maurer2025real} implemented a hardware-friendly approximate XZ-Relay-BP decoder on an AMD VU19P FPGA. They replaced floating point multipliers by bit-shifters and adders, with quantization down to 4 bits and limiting memory factors to a predetermined range. This allows them to expand BP in a fully unrolled fashion to meet the real-time constraint at the cost of slightly higher LER. However, the internal state update mechanism enforces that each Variable Node Unit (VNU) must carry its own RNG, as well as required auxiliary logic to blend the bias (which consists of multipliers and adders). As a result, their VNUs collectively take up $67.2\%$ of LUTs. For full XYZ decoding however, there are substantially more (71280) variable nodes than split X/Z (8784), and XYZ-Relay-BP requires at least 6 bits of precision to retain its LER\cite{maurer2025real}. This makes an unrolled XYZ-Relay-BP architecture, similar to that implemented for the XZ-Relay-BP, challenging to fit onto a modern FPGA without significant modifications.

Because qDither's RSF design eliminates the need for per-iteration bias blending, we can replace the RNG and auxiliary logic with simpler oscillation detection hardware. The simplest implementation can include an XOR gate between the most significant register bits of the previous and current marginals, a multiplexer, and an adder. Furthermore, RIS blending only occurs during posterior probability hand-off between attempts and requires additional logic at the decoder level. This permits a much lighter VNU design, making qDither a viable candidate for scaling to full XYZ decoding.

\section{Conclusion}
In this work, we showed that the internal state evolution from Relay-BP is not a necessary condition for belief propagation to escape non-converging behaviors. RIS allows the decoder to start its additional attempt from an updated prior, and RSF enables oscillation dampening and faster convergence. Based on our experiments, we achieve near or better than Relay-BP performance on surface and bivariate-bicycle codes with projected less hardware overhead. QDither trades a moderate increase in additions for a reduction in area through the elimination of node-level RNG and a 98\% reduction in multiplications. Our future directions include hardware architectures, efficient hyperparameter tuning, and benchmarking on additional code families.

\bibliographystyle{IEEEtran}
\bibliography{AutoencoderProject}

\end{document}